# 3D structure and stability prediction of DNA with multi-way junctions in ionic solutions

Xunxun Wang[1], Ya-Zhou Shi[2,*]

[1] Guizhou Key Laboratory of Microbio and Infectious Disease Prevention & Control, School of Biology and Engineering, Guizhou Medical University, Guiyang 550025, China

[2] Research Center of Nonlinear Science and School of Mathematical and Physical Sciences, Wuhan Textile University, Wuhan 430073, China

## Abstract

Understanding the three-dimensional (3D) structure and stability of DNA is fundamental for its biological function and the design of novel drugs. In this study, we introduce an improved coarse-grained (CG) model, incorporating a more refined electrostatic energy term, the replica-exchange Monte Carlo algorithm, and the weighted histogram analysis method. The enhanced model predicts the 3D structures and stability of DNA with multi-way junctions (three-way and four-way) in various ionic environments, going beyond traditional single-stranded DNA (ssDNA) and double-stranded DNA (dsDNA). Our model demonstrates remarkable accuracy in predicting the structures of DNAs with multi-way junctions from sequences and offers reliable estimates of their thermal stability across a range of sequences and lengths, with both monovalent and divalent salts. Notably, our analysis of the thermally unfolding pathways reveals that the stability of DNA with multi-way junctions is strongly influenced by the relative stabilities of their unfolded intermediate states, providing key insights into DNA structure-function relationships.

---

*To whom correspondence should be addressed: yzshi@wtu.edu.cn.

**Introduction**

DNA is a fundamental macromolecule in living organisms, playing pivotal roles in functions such as genetic information storage and transmission [1], protein synthesis regulation [2], and gene expression control [3]. Its multifunctionality underpins various nano-biomedical and technological applications [4]. Moreover, DNA must adopt particular 3D conformations to perform specific functions [5,6]. For instance, the faithful transmission of genetic information of DNA relies on its dynamic 3D structure changes [7]. Additionally, enhancers and promoters in DNA sequences interact through the DNA 3D folding to modulate the binding of transcription factors, thereby initiating or repressing gene expression [8]. Moreover, DNA with multi-way junctions exhibits greater structural complexity than simple ssDNA or dsDNA forms [9], potentially playing a crucial role in biological processes like gene regulation and molecular recognition [10]. Hence, a more profound understanding of these complex structures is essential to elucidate diverse biological functions of DNA fully.

The 3D structures of DNAs can be experimentally determined through X-ray crystallography, NMR spectroscopy, and cryo-electron microscopy [11,12]. However, due to the high cost of these experimental techniques, the number of DNA 3D structures deposited in the Protein Data Bank (PDB) remains limited [11,13], especially when compared to the vast number of DNA sequences in GenBank [14,15]. To complement experimental approaches, various theoretical and computational models have been developed to predict DNA 3D structures. These models can be broadly classified into three categories. The first category includes template-based fragment assembly models, such as 3dRNA/DNA [16], which predict DNA 3D structures based on secondary structure. The success of the model depends on the quality and diversity of the template library. The second category includes deep learning-based models, such as AlphaFold3 [17,18], which leverage deep learning techniques to predict DNA 3D structures from established structure datasets. However, the diversity of DNA structures can limit the accuracy of the model in predicting DNA 3D structures. The third category involves physics-based models, such as all-atom molecules dynamics (e.g., Amber [19,20], and Charm [21,22]), Scheraga's model [23], NARES-2P [24], which utilize specific force fields combined with conformational sampling algorithms like Monte Carlo/Replica Exchange Monte Carlo (MC/REMC) [25,26] or Molecular Dynamics/Replica Exchange Molecular Dynamics (MD/REMD) [27-29]. These models can solely predict DNA 3D structures from the sequence. However, despite simplifying nucleotides or utilizing efficient sampling algorithms, these models still face challenges in accurately predicting the 3D structures of complex DNA molecules based solely on sequence information, particularly in terms of multi-way junctions (≥three-way junctions) [30-32].

Due to its polyanionic nature, the folding and dynamics of DNAs are heavily influenced by metal ions (e.g., $Na^+$ and $Mg^{2+}$) in solution. Several existing models, including 3SPN, oxDNA, TIS, and NARES-2P, have incorporated electrostatic interactions using the Debye-Hückel approximation or mean-field multipole–multipole potentials [33-37]. These models have successfully captured certain monovalent salt-dependent properties, such as torsional stiffness and melting temperatures. However, accurately predicting the 3D structure and thermal stability of DNA, particularly in the presence of monovalent/divalent ions, remains a challenging task [13]. Moreover, these models have not yet fully addressed the folding problem of complex DNA, especially molecules with intricate spatial structures, such as DNA with multi-way junctions (e.g., three-way and four-way junctions) in the physiological environment [38]. Recently, we proposed a three-bead CG model to simulate DNA folding solely from the sequence, incorporating an implicit electrostatic potential [13]. This model has been proven effective in predicting the 3D structures and stability of ssDNA and dsDNA in ion solutions. However, our previous model based on a traditional MC annealing algorithm has significant limitations [13]. Its efficiency in exploring the conformational space is particularly inadequate for simulating DNA structures with intricate multi-way junctions.

In this work, we have carefully refined our previously developed CG model by incorporating a refined electrostatic potential, a new sampling algorithm REMC, and along with the weighted histogram analysis method (WHAM). Specifically, the new model includes: (i) a structure-based electrostatic energy term to capture electrostatic interactions between DNA and monovalent/divalent ions (ii) the use of a more efficient REMC instead of the MC simulated annealing algorithm [13,39]; (iii) WHAM to analyze the thermal stability of DNA; and (iv) an all-atom rebuilding algorithm of the identified CG structures. We then applied the model to predict the 3D structures and stability of DNA with multi-way junctions extending beyond minimal ssDNA and dsDNA in both monovalent and divalent ion solutions. Finally, the model was used to analyze the thermally unfolding pathways of DNA with multi-way junctions.

## Materials and methods

### Coarse-grained representation for DNA structures

In our present CG model, we extend our previous CG model, where each DNA nucleotide is represented by a set of three CG beads [13]. These beads correspond to the phosphate (P) atom, the C4′ atom, and either the N1 atom in pyrimidines or the N9 atom in purines (Fig. 1A). This CG approach effectively captures the key structural features of DNA, including the backbone and base pairs [40]. The P, C, and N beads are modeled as spheres with van der Waals radii of 1.9 Å, 1.7 Å, and 2.2 Å, respectively, with a unit negative charge assigned to the P bead.

**Coarse-grained force field**

In our present CG model, the whole energy potential for an DNA chain is given by:

$$U = U_b + U_a + U_d + U_{exc} + U_{bp} + U_{bs} + U_{cs} + U_{el}, \qquad (1)$$

the bonded interaction energies in the DNA chain are represented by $U_b$, $U_a$, and $U_d$, corresponding to bonded length, bonded angle, and bonded dihedral energies, respectively. The remaining terms in Eq. 1 describe non-bonded interactions. $U_{exc}$ captures repulsive forces due to the excluded volume between CG beads. $U_{bp}$, $U_{bs}$, and $U_{cs}$ represent the interactions between base pairs, base stacking, and coaxial stacking, respectively. The final term, $U_{el}$ accounts for electrostatic interactions between phosphate groups along the DNA chain, implicitly incorporating the effects of both monovalent and divalent ions. Notably, $U_{el}$ is seldom included in other DNA 3D structure prediction models.

Our previous CG model successfully predicted the 3D structures of relatively simple DNA topologies, such as ssDNA hairpins, dsDNA duplex/kissing complexes, and minimal H-type pseudoknots [13], but it struggled with more complex structures, particularly DNA with multi-way junctions (≥ three-way junctions). This limitation was likely due to the MC annealing algorithm used in the previous model, which resulted in inadequate conformational sampling. DNA with multi-way junctions exhibits greater complexity and a more rugged potential energy landscape in both its secondary and 3D structures, posing significant challenges for sampling algorithms. To address this, the new model incorporates a REMC algorithm, which enhances sampling efficiency by enabling exchanges between adjacent replicas at different temperatures. This sampling algorithm helps the system escape local energy minima and facilitates a more thorough exploration of conformational space, improving prediction accuracy for complex DNA structures. Furthermore, to calculate the thermal stability of DNA based on the predicted structure trajectories, we applied the WHAM method to eliminate bias ($E$ and $T$) and accurately obtain the thermodynamic properties of the canonical ensemble [41]. Additionally, based on RNA 3D structure model from the Tan group [40,42,43], an implicit structure-based electrostatic potential has been incorporated into the present model to more accurately capture the interactions between DNA and ions, it implicitly accounts for the effects of monovalent and divalent ions in $U_{el}$ through the combination of counterion condensation theory [44] and the tightly bound ion (TBI) model [45,46], the more detailed describe of electrostatic potential can be found in Supporting Information. Consequently, the model can simulate DNA folding in solutions containing mixed monovalent and divalent ions. Meanwhile, the other energy terms in the force field remain consistent with the previous version [13]. Two sets of bonded potentials Para$_{loop}$ (for single-stranded regions) and Para$_{helix}$ (for stem/helix regions) were used to describe the backbone of DNA. Detailed descriptions of the force field and its parameters can be found in the Section "Force Field Description", and Tables S2 & S3 in the Supporting Information.

**Replica-exchange Monte Carlo simulations**

In our present CG model, to predict the structures of DNA with multi-way junctions, which typically involve a rugged energy landscape, we employed the REMC algorithm [25] for conformational sampling. The REMC algorithm allows DNA conformations of adjacent temperature that are trapped in local energy minima at low temperatures to escape to those at higher temperatures [47], enabling the system to explore a wider range of conformations with lower energies and thus fold into near-native states with higher probability compared to the conventional MC annealing algorithm. Specifically, the REMC algorithm in our present model involves 10 replicas at temperatures of 25°C, 31°C, 37°C, 45°C, 54°C, 64°C, 74°C, 86°C, 98°C, and 110°C. The same initial DNA conformation is simulated in parallel at these 10 temperatures. Conformational changes for each replica are made using pivot moves according to the Metropolis algorithm [26]. At periodic intervals, replica exchanges occur between adjacent temperatures, with an exchange probability $p$=1 if $\Delta = (1/k_B T_i - 1/k_B T_j)(E_j - E_i) \leq 0$, and $p=\exp(-\Delta)$, if $\Delta > 0$, where $E_i$ and $E_j$ represent the energies of the two replicas at temperatures $T_i$ and $T_j$ [47].

**Identifying top-scored coarse-grained 3D structures**

In our present CG model, to identify the top-scoring structures from the predicted CG 3D structure ensemble, we first used our CG energy function to select the 1000 DNA conformations with the lowest energy from the predicted ensemble at the lowest temperature (25°C). Next, we applied clustering algorithms to group similar structures by calculating the RMSD values between all pairs of structures within the ensemble. The cluster with the largest number of structures, within a predefined RMSD threshold was identified, and its members were removed from the initial set. This process was repeated iteratively until all structures were assigned to a cluster. Based on our experience and RNA 3D structure prediction model SimRNA [48], we typically used a clustering threshold of 0.1 Å times the sequence length (e.g., 5 Å for a 50-residue sequence). Finally, the medoids of the three largest clusters with the decoy with the lowest energy were selected. This procedure was followed in the current study.

**Rebuilding all-atom structure**

In our present CG model, the top-scoring CG structures (top-1 or top-n) are converted into all-atom structures for practical application. First, we constructed a small library of five non-redundant all-atom structures for each type of base pairing (G-C, A-T, C-G, T-A) and single nucleotides (A, G, C, T) by clustering according to their mutual RMSDs. Next, based on the sequence and secondary structure of the top-scoring CG structures, each all-atom base pair or nucleotide is aligned to the corresponding CG stem or loop region by superimposing their respective atoms onto the top-scored CG structures. The base-pairing or single nucleotide with the smallest RMSD to the CG atoms of the top-scored CG structure is retained for each corresponding base pair or single nucleotide. This process is repeated along the entire DNA chain to construct the top-scored all-atom 3D DNA structure, and the detailed rebuilding progress can be seen in Fig. S1. Finally, to eliminate potential steric clashes and chain breaks in the rebuilt all-atom structures, a structure refinement step is performed using the QRNAS method [49].

**Analyzing structure stability with weighted histogram analysis method**

In our present CG model, we applied REMC simulations to assess the thermal stability of DNA molecules by calculating the population fractions of different structural states at various temperatures using the WHAM [50]. This methodology allowed us to analyze key thermal stability characteristics, including melting temperatures and unfolding pathways of DNAs. The process began with the generation of REMC trajectories for the DNA across a range of temperatures. We then discretized the relevant reaction coordinates structural states ($S$) and energies ($E$) into bins, where $S_j$ ($j = 1, 2, \ldots, F$) represents structural states and $E_k$ ($k = 1, 2, \ldots, 100$) represents energy levels, with each set ($j, k$) corresponding to a small state in the WHAM. Finally, we computed the probability $p^{\circ}_{(j,k)}$ for each small state at the temperature of interest $T$ through iterative equations [41].

$$p^{\circ}_{(j,k)} = \frac{\sum_{i=1}^{M} n_{i,(j,k)}}{\sum_{i=1}^{M} N_i Z_i c_{i,(j,k)}}; \quad (2)$$

$$Z_i^{-1} = \sum_{j,k} c_{i,(j,k)} p^{\circ}_{(j,k)}. \quad (3)$$

Here, $M$ denotes the number of replicas, while $n_{i,(j,k)}$ represents the number of occurrences of the small state ($j, k$) at the $i$-th temperature. $N_i$ is the total number of conformations at the $i$-th temperature, and $c_{i,(j,k)}$ is the temperature bias factor. Finally, $f_{S_j}(T)$ represents the relative fraction of each structural state of the DNA at temperature $T$, and was calculated using the following equation:

$$f_{S_j}(T) = \sum_{k=1}^{100} p^{\circ}_{(j,k)}, \quad (4)$$

where $S_j$ represents various structural states of the DNA, such as folded (F), unfolded (U), and intermediate (I) conformations. The fraction $f_{S_j}(T)$ of each state at temperature $T$ provides insight into the thermal stability and thermally unfolding pathway of DNAs. To quantify the melting temperature, the fractions of the folded state, $f_F(T)$ (where $F = S_F$) and the unfolded state $f_U(T)$ (where $U = S_1$) are fitted to a two-state model (Eqs. 5 and 6). This allows for a more accurate assessment of the thermal stability and transition behavior of the DNA.

$$f_F(T) = \frac{1}{1+e^{(T-T_{m1})/dT_1}}; \tag{5}$$

$$f_U(T) = 1 - \frac{1}{1+e^{(T-T_{m2})/dT_2}}. \tag{6}$$

Here, $T_{m1}$ and $T_{m2}$ correspond to the melting temperatures for the transitions from folded to intermediate (F→I) and from intermediate to unfolded (I→U), respectively. For a comprehensive explanation of the WHAM method and a detailed definition of the different structural states (folded, F; unfolded, U; and intermediate, I) used in this study, please refer to the Supporting Information.

## Results and discussion

In this work, we applied our newly developed CG model to predict the 3D structures of complex DNA molecules, including those with three-way and four-way junctions, extending beyond the simpler ssDNA and dsDNA structures. We then evaluated the thermal stability of these complex DNA structures in both monovalent and divalent ion solutions, surpassing the simple ssDNA, dsDNA, and pseudoknot structures explored in our previous work [13]. Finally, we performed a comprehensive analysis of the thermally unfolding pathways for DNA molecules containing three-way and four-way junctions.

### Predicting 3D structures of DNA with multi-way junctions

The present model predicts the 3D structures of DNA through a series of steps. First, an initial random conformation of the DNA chain is generated based on the bonded potential $U_{bond}$ and the excluded-volume potential $U_{exc}$ (Eq. 1). Using this initial conformation, a REMC simulation is performed to sample the conformation space of DNA. During this simulation, the Para$_{loop}$ bonded potential parameters are applied to the entire DNA chain. In the second step, since REMC produces an ensemble of candidate structures, we employ the energy function in the CG model and a clustering algorithm [51] to select the top scoring-structures from the predicted structure ensemble, typically at the lowest temperature (e,g., selecting from 1000 candidates at 25°C). The third step involves structure refinement: we perform a MC simulation at 25°C on the top structure from the folding process. In this step, two distinct sets of bonded potential parameters Para$_{loop}$ for the loop regions and Para$_{helix}$ for the helix base-pairing regions are used to better capture the helix geometry of the stems [52]. In this work, to evaluate the performance of our model, we focus on the top structures based on RMSD metrics (top-1 RMSD and RMSD$_{min}$). The RMSD provides a measure of the global deviation of the predicted structures from the native ones [53,54]. Additionally, to complement the RMSD analysis, we also utilized the F1-score ($F1 = 2 \times PR \times SN/(PR + SN)$) to assess the accuracy of the predicted secondary structures of DNAs, as described in [55,56]. The F1-score evaluates the consistency of base-pairing interactions between the predicted and native structures, where precision (PR) is calculated as $PR = TP/(TP + FP)$ and sensitivity as $SN = \frac{TP}{TP+FN}$, with TP, FP, and FN denoting true positives, false positives, and false negatives, respectively.

**Structure prediction for DNA with multi-way junctions and comparison with other models**

In this work, we used 4 DNAs with multi-way (3 three-way and 1 four-way) junctions to evaluate the present model on predicting structures of complex DNA. Since the experimental structures of some of these DNAs were from X-ray crystallography, here we only predicted 3D structures of the 4 DNAs at 1 M [Na$^+$], a standard salt condition [57-59].

Fig. 2 presents both the secondary and 3D structures with the minimum RMSDs and top-1 RMSDs from the predicted ensembles of 4 complex DNAs, comparing them with the native structures. This comparison demonstrates that the present model successfully captures the 3D topologies of complex DNA solely from the sequences. In Figs. 2A-D, we showed the RMSD values of the structures that most closely resemble the native conformations within the structure ensembles of model for all 4 complex DNAs. These RMSD values range from ~5 Å to ~8 Å, indicating that our model can predict near-native 3D structures for complex DNAs based solely on their sequences. Additionally, we provided the RMSDs of the top-1 structures in Fig. 2, where the RMSDs of 4 DNAs are <10 Å, demonstrating the ability of model to generate reasonably accurate predictions.

To further evaluate our present model, we compared it with two top methods for DNA 3D structure prediction: 3dRNA/DNA [16] and AlphaFold3 [17]. In the comparison, we used only the DNA sequences as input for both models, as our model relies solely on sequence information for structure predictions, and for 3dRNA/DNA and Alphafold3, we used their webservers to predict the DNA 3D structure. Initially, we applied 3dRNA/DNA to predict the 3D structures of the 4 complex DNAs, using the default 'RNAfold' settings for secondary structure prediction. As shown in Fig. 3C, the average RMSDs of the top-1 structures predicted by 3dRNA/DNA for the 4 complex DNAs were ~15.6 Å for the 'RNAfold' and 'Optimization' settings. These values are notably larger than the ~8.8 Å RMSD obtained from our model. Next, we used AlphaFold3 to predict the 3D structures of the same four complex DNAs. As shown in Fig. 3B, the average RMSD of the top-1 structures for these DNAs predicted by AlphaFold3 was ~9.2 Å, which are larger than those obtained by our model.

Furthermore, as illustrated in Fig. 3C, the mean F1-score of the top-1 structures predicted by our model for the 4 complex DNAs is ~0.88, which exceeds the F1-score obtained from 3dRNA/DNA (~0.25), but is slightly lower than the value from AlphaFold3 (0.93). Although our model shows a slight gap compared to deep learning-based method, it outperforms traditional models like 3dRNA/DNA. Additionally, the average minimum F1-score of our predictions for the 4 complex DNAs is relatively large (0.93), suggesting that incorporating a more reliable structural refinement step or a scoring function could further improve the prediction accuracy of our model.

**Predicting stabilities of DNA with multi-way junctions in monovalent/divalent ion solutions**

The functionality of DNA is influenced not only by its static structure but also by its stability under different conditions. In comparison to other DNA 3D structure prediction methods, our model offers the unique capability to predict the thermal stability of complex DNAs (including those containing three-way and four-way junctions) in solutions with both monovalent and divalent ions. To assess the effectiveness of model in predicting thermal stability, we applied it to 6 diverse DNA sequences (3 three-way junctions and 3 four-way junctions), with detailed information of 6 DNAs can be seen Fig. S2 in the Supporting Information. As outlined in the Materials and Methods section, our model generates the fractions of folded, unfolded, and intermediate states, which are then used to analyze the unfolding pathways of DNA structures with multi-way junctions. Further details regarding the WHAM used to process these results can be found in the Supporting Information.

*Predicting thermal stabilities of DNA with multi-way junctions*

As shown in Fig. 4, to demonstrate the use of our model in predicting the thermal stability of DNA with multi-way junctions, we used a DNA with three-way junction (3WJ) as the example (sequence: 5′-GAAATTGCGCTTTTTGCGCGTGCTTTTTGCACAATTTC-3′) [60]. For the 3WJ, there are eight distinct structural states, including the fully folded state (F=$S_8$) where all three stems are retained, several intermediate states (with one or two stems retained), and the unfolded state (U = $S_1$) represents the conformation where all three stems are resolved (see Fig. S3 in the Supporting Information). By applying

Eqs. 2 and 3, we calculated the probabilities $p^{\circ}_{(j,k)}$ for each small state and subsequently determined the fractions of these states using Eq. 4. The melting temperatures ($T_{m1}$ and $T_{m2}$) are then derived by fitting the fractions of folded and unfolded states ($f_F$ and $f_U$, respectively) to a two-state model (Eqs. 5 and 6). According to our model, the predicted melting temperatures for the 3WJ are $T_{m1}$= 47.6°C (transition from folded state to intermediate state) and $T_{m2}$ = 79.5°C (transition from intermediate state to unfolded state). It is important to note that due to the application of a two-state model, our present model can only provide the overall melting temperatures, rather than the melting temperature for each stem. Additionally, for the more complex secondary and 3D structure DNA with four-way junctions 4WJ (sequence: 5′-GAAATTGCGCTTTTTGCGCATATCTTTTTGATAGGTGCTTTTTGCACAATTTC-3′) [60], as illustrated in Figs. 4D-E, the melting temperatures predicted by our model are $T_{m1}$ = 38.4°C and $T_{m2}$ = 82.5°C, calculated using the same methodology as for the 3WJ.

Furthermore, we made predictions on the melting temperature for the other 4 DNAs (L-3WJ, R-3WJ, L-4WJ, and R-4WJ) with different sequences [60]. The sequences and secondary structures of the 4 DNAs can be found in Fig. S2 in the Supporting Information. As shown in Table 1, for 3 different DNAs with three-way junctions (3WJ, L-3WJ, and R-3WJ), the predicted melting temperatures ($T_{m1}$ and $T_{m2}$) by the present model are in good agreement with experimental data, and the mean deviations between the predictions and experimental data are ∼2.5℃ for $T_{m1}$ and ∼4.1℃ for $T_{m2}$, indicating reliable predictions on the thermal stability of DNA with multi-way junctions [60]. Additionally, for R-3WJ, the melting temperature $T_{m1}$ of experimental is 54.6℃, while the $T_{m1}$ of our model is 50.6℃, our prediction is significantly lower than experiment. This is because the melting temperature $T_{m1}$ of the experimental value refers to the melting of two stems, while the $T_{m1}$ predicted by our model refers to the melting of single stem. To further examine the performance of our model, we roughly treated the intermediate state containing two stems as a folded state, so the predicted $T_{m1}$ of our model is 55.7℃, quite close to the experimental values. Overall, our predicted melting temperatures for DNA with three-way junctions are highly consistent with the experimental results. For 4WJ, Similar to 3WJ, there is also a lack of $T_{m1}$ (one stem resolved) and $T_{m2}$ (transition from an intermediate state containing one stem to a fully unfolded state) for 4WJ in the experiment. The melting temperature $T_{m1}$ predicted by our model is all lower than the experimental value, while the melting temperature $T_{m2}$ is almost higher than the experimental value because the melting temperature measured in the experiment is greater than the melting of one stem instead of a single stem in our model. But this also indirectly proves that the melting temperature predicted by our model is reasonable. Overall, our model can reliably predict the thermal stability of DNA with multi-way junctions in different sequences.

*Effect of monovalent ion on DNA with multi-way stability*

DNA 3D structures and stabilities are highly sensitive to the ionic conditions of the solution due to their polyanionic nature [61-63]. In this work, we examined the monovalent salt ($Na^+$) dependence of complex DNA stability, using a three-way junctions (3WJ) and a four-way junctions (4WJ) as paradigms [64]. As shown in Fig. 5, the melting temperatures ($T_{m1}$ 35.2°C/39.8°C and $T_{m2}$ 65.2°C/71.1°C) for 3WJ predicted by our model are in good agreement with experimental values ($T_{m1}$ 33.6°C/36.1°C and $T_{m2}$ 70.3°C/72.5°C) at [$Na^+$] of 100 mM/200 mM [64]. Additionally, as shown in Fig. 5, we also predicted the melting temperatures for 3WJ and 4WJ over a wide range of [$Na^+$]. Fig. 5A demonstrates that the stabilities of both the DNA and intermediate hairpin structures increase with increasing [$Na^+$], which is attributed to the stronger ion neutralization effect at higher [$Na^+$], particularly for compactly folded or intermediate state structures. Notably, $T_{m1}$ corresponding to the transition from the folded state to the intermediate hairpin state, increases less sharply with [$Na^+$] than $T_{m2}$, which represents the transition from the intermediate hairpin state to the unfolded state. This is because the intermediate hairpin (containing single stem) structures involves greater charge accumulation compared to folded state (containing three stems). Here we separately calculated the average rotation radius (Rg) of all conformation of the folded state and intermediate states at 25℃. For 3WJ, we found that the average Rg of the two intermediate states i2 (15.0 Å) and i2′ (15.9 Å) are smaller than that of the folded state F, indicating that the structural compactness of the intermediate state is greater than that of the folded state (18.4 Å). For 4WJ, the three intermediate states-i3 (11.0 Å), i3′ (10.5 Å), and i3″ (10.5 Å)-display smaller average Rg values compared to the folded state F (21.9 Å). This suggests that the intermediate states are more compact in structure than the fully folded state; see Fig. S4. Therefore, the stability of intermediate hairpin state is generally more dependent on ion concentration than fully folded states.

*Effect of divalent ion on DNA with multi-way junctions stability*

As shown in Fig. 5, the predicted melting temperature ($T_{m1}$ and $T_{m2}$) was analyzed across a wide range of [$Mg^{2+}$], with [$Na^+$] fixed at 10 mM (Fig. 5B) and 100 mM (Fig. 5C), respectively. The model successfully captures the competitive effects of $Na^+$ and $Mg^{2+}$ on the stability of DNA with three-way junctions (3WJ). At low [$Mg^{2+}$] (e.g., 1 mM), the stability of 3WJ is primarily governed by the background $Na^+$, and the $T_m$ values ($T_{m1}$ and $T_{m2}$) closely resemble those observed in pure $Na^+$ solutions. As [$Mg^{2+}$] increases (~>1 mM), the stability of the 3WJ structure is significantly enhanced due to the stronger binding affinity of $Mg^{2+}$ especially for $T_{m2}$. This effect reaches a saturation point at higher [$Mg^{2+}$] ( ~>100 mM) concentrations, which is attributed to ion binding saturation. This is due to the anti-cooperative binding between $Na^+$ and $Mg^{2+}$, coupled with the more efficient stabilization provided by $Mg^{2+}$. Additionally, as shown in Figs. 5E and 5F, a similar behavior is observed for four-way junctions (4WJ). Overall, the model demonstrates its ability to accurately predict the folding behavior of DNA with multi-way junctions in the presence of divalent ions.

**Thermally unfolding pathway of DNA with multi-way junctions**

The metastable structures of DNA are crucial to their functions [65], and understanding the thermally unfolding pathways of DNA with multi-way junctions is essential for exploring their functional mechanisms [66]. The present model effectively captures the melting temperatures of DNA with three-way and four-way junctions and can also analyze the thermally unfolding pathways of complex DNA structures, extending beyond the minimal ssDNA and dsDNA, and H-type pseudoknots covered in our previous model [13,40,42]. By performing REMC simulations for each DNA structure under specified solution conditions, we can calculate the fractions of folded (F), unfolded (U), and intermediate states (I) at different temperatures, as detailed in the Materials and Methods. In this work, we focus on two typical DNA with multi-way junctions structures 3WJ (DNA with three-way junction) and 4WJ (DNA with four-way junction) to analyze their unfolding pathways beyond the minimal ssDNA and dsDNA [13]. Here, the intermediate states are denoted as I1, I2, ..., corresponding to the number of melting stems, while states without labels or those labeled with "′", "″", "‴", etc., represent states with the highest, second-highest, third-highest fraction, and so on.

*3WJ: a DNA with three-way junctions*

We thoroughly analyzed the thermally unfolding pathway of the 3WJ at 1M NaCl and identified eight key structural states, and the fractions of these major structural states were tracked as a function of temperature. As shown in Fig. 6A, which illustrates how the fractions of the folded (F), unfolded (U), and major intermediate states (I1, I2, and I2′) change with temperature. At temperatures < ~30°C, the 3WJ predominantly remains in the F. As the temperature increases from ~30°C to ~50°C, the fraction of the folded state F decreases from ~100% to ~32%, while the fractions of intermediate states I1, I2, and I2′ gradually increase to ~32%, ~32%, and ~5%, respectively. Between ~50°C and ~70°C, the fractions of the F and I1 drop to ~10%, while the fractions of I2 and I2′ rise, reaching their peak values of ~55% and ~10%, respectively. At ~110°C, the system transitions into the unfolded state U. Based on these observations, we inferred the primary unfolding pathway for 3WJ as F→I1→I2→U, with the fraction of I2 being significantly higher than that of I1 and I2′ at ~70°C. Additionally, two minor pathways were observed: F→I2→U and F→I1→I2′→U, with the former pathway having a slightly higher flux than the latter.

To investigate the unfolding mechanism of the 3WJ, we computed the free energies of all possible intermediate states using the nearest-neighbor model with Mfold parameters [67]. Here, we used the webserver of Mfold to calculate the free energy of intermediate states, and the eight structural states can be found in Fig. S3. Initially, we focused on the intermediate states with a single stem melted, specifically the I1 state (stem 1 melted), I1′ state (stem 2 melted), and I1″ state (stem 3 melted) at 1 M [Na$^+$]. The relative free energy between the I1 and I1′ states ($\Delta\Delta G_{I1,I1'} = \Delta G_{I1} - \Delta G_{I1'}$) was ~-1.2 kcal/mol at ~50°C, and between the I1 and I1″ states ($\Delta\Delta G_{I1,I1''} = \Delta G_{I1} - \Delta G_{I1''}$) was ~-2.8 kcal/mol at ~50°C. These findings indicate that the I1′ and I1″ states are significantly less stable than the I1 state, which is consistent with the higher population of the I1 state relative to the I1′ and I1″ states at ~50°C. We then examined the free energy of the I2 state, where stems 1 and 3 are melted. Below ~50°C, the free energy of the I2 state is higher than that of the I1 state, but at temperatures above ~70°C, the I2 state becomes more stable than the I1 state. Specifically, $\Delta\Delta G_{I2,I1} = \Delta G_{I2} - \Delta G_{I1}$ was ~0.5 kcal/mol at 50°C and -1.3 kcal/mol at 70°C, suggesting that the I2 state is less populated at lower temperatures and becomes more abundant at elevated temperatures, supporting the major unfolding pathway F→I1→I2→U with increasing temperature. Finally, we calculated the free energies of intermediate states with two stems melted, such as I2′ (stems 1 and 2 melted) and I2″ (stems 2 and 3 melted), which have negligible populations. The I2″ state exhibited significantly higher free energy than both the I2 and I2′ states, consistent with its negligible fraction. These results indicate that the unfolding of 3WJ primarily follows the major pathway F→I1→I2→U, with two minor pathways, F→I2→U and F→I1→I2′→U, also contributing. The relative stability of the I2 state, combined with the higher stability of the I1 and I1′ states compared to other intermediates, explains the observed unfolding behavior.

*4WJ: a DNA with four-way junctions*

In addition to analyzing the thermal unfolding pathway of DNA with three-way junctions, we also investigated the unfolding behavior of DNA with four-way junctions (4WJ) at 1 M NaCl. The intermediate states of 4WJ are much more complex than that of 3WJ, exhibiting sixteen distinct structural states (see Fig. S5). These states were analyzed by calculating the fractions of each state at various temperatures. As shown in Fig. 7A, the fractions of the folded state (F), the fully unfolded state (U), and the major intermediate states are plotted as functions of temperature. Notable intermediate states include I1 (with Stem 3 melted), I1′ (with Stem 1 melted), I2 (with both Stem 1 and Stem 3 melted), and I3 (with all three stems melted), as illustrated in Fig. 7E. Below ~20°C, the 4WJ remains predominantly in the folded state, with all four stems intact. As the temperature rises from ~20°C to ~40°C, the fraction of the folded state decreases from ~100% to ~25%, while the fractions of intermediate states I1, I1′, and I2 gradually increase to ~20%, ~15%, and ~40%, respectively. At ~50°C, the fractions of the folded state and the intermediate states I1 and I1′ decline to ~0%, and the fraction of I2 reaches its maximum at ~55%. As the temperature increases to ~70°C, the fraction of I2 decreases to ~15%, while the fraction of I3 increases to ~55%. As shown in Fig. 7A, at ~120°C, the system reaches the fully unfolded state (U). Based on these observations, the thermal unfolding pathway of the 4WJ is proposed (Fig. 7B). The primary unfolding pathway follows F→I2→I3→U, as the fraction of I2 is significantly larger than that of I1 and I1′ at ~50°C. Four minor unfolding pathways are also observed: F→I1→I2→I3→U, F→I1′→I2→I3→U, F→I1→I2→I3→U, and F→I1′→I2→I3→U, with the first pathway displaying slightly greater flux than the others.

To explore the unfolding pathway of the four-way junction (4WJ), we calculated the free energies of all possible intermediate states using the nearest-neighbor model with Mfold parameters [67]. To account for the complexity of the intermediate state structures, we modified the loop sequences by substituting them with 'X' to ensure correct secondary structure prediction by Mfold. Initially, we calculated the free energies of intermediate states with a single stem melted, including I1 (Stem 3 melted), I1′ (Stem 1 melted), I1″ (Stem 4 melted), and I1‴ (Stem 2 melted) at 1M NaCl. The relative free energy between I1 and I1′ ($\Delta\Delta G_{I1,I1'} = \Delta G_{I1} - \Delta G_{I1'}$) was ~-0.4 kcal/mol at ~30°C, while the relative free energy between I1 and I1″ ($\Delta\Delta G_{I1,I1''} = \Delta G_{I1} - \Delta G_{I1''}$) was -~1.0 kcal/mol, and between I1 and I1‴ ($\Delta\Delta G_{I1,I1'''} = \Delta G_{I1} - \Delta G_{I1'''}$) was -~2.5 kcal/mol. Additionally, the relative free energy between I1′ and I1″ ($\Delta\Delta G_{I1',I1''} = \Delta G_{I1'} - \Delta G_{I1''}$) was -~0.6 kcal/mol, and between I1′ and I1‴ ($\Delta\Delta G_{I1',I1'''} = \Delta G_{I1'} - \Delta G_{I1'''}$) was ~-2.2 kcal/mol at ~30°C. These results indicate that the I1″ and I1‴ states are significantly less stable than I1 and I1′, which is consistent with the higher population of I1 and I1′ relative to I1″ and I1‴ at ~30°C. We then examined the free energy of the I2 state (Stem 1 and Stem 3 melted). At lower temperatures (< ~40°C), the free energy of I2 was higher than that of I1 and I1′, but at higher temperatures (>~50°C), it became more stable. Specifically, at ~30°C, $\Delta\Delta G_{I2,I1}$ was ~1.8 kcal/mol and $\Delta\Delta G_{I2,I1'}$ was ~1.4 kcal/mol, while at ~50°C both were ~-0.6 kcal/mol. These results suggest that the I2 state is less populated at low temperatures but becomes the dominant state as the temperature increases, supporting the major unfolding pathway F→I2→I3→U. The minor pathways F→I1→I2→I3→U and F→I1′→I2→I3→U were also observed, and with a slightly higher flux through the F→I2→I3 →U pathway. Additionally, we also investigated intermediate states with two stems melted, such as I2′ (Stems 1 and 3 melted) and I2″ (Stems 1 and 4 melted), which showed negligible contributions to the unfolding process. These states exhibited higher free energies than I2, further supporting their minimal role. Finally, we considered states with three stems melted, including I3′ (Stems 1, 2, and 3 melted), I3″ (Stems 1, 2, and 4 melted), and I3‴ (Stems 2, 3, and 4 melted), which also had negligible fractions, with the I3 state being the most populated. Therefore, the unfolding of 4WJ follows a major pathway, F→I2→I3→U, with minor pathways F→I1→I2→I3→U, F→I1′→I2→I3→U, and F→I1→I2→I3→U, all involving the relatively stable I2 state and the less stable I1 and I1′ states, which are more stable than other intermediate states.

## Conclusion

In conclusion, this study represents a significant advancement in the prediction of DNA 3D structures and their thermal stability, with a particular focus on multi-way junctions. By refining our CG model and integrating the REMC algorithm, we have expanded the applicability of the model to complex DNA structures, including three-way and four-way junctions, under both monovalent and divalent ion conditions. These improvements enable a more precise investigation of the thermal unfolding behaviors of DNA junctions, offering valuable insights into their stability and conformational dynamics. The main contributions of this work are as follows:

1. Accurate Structure Predictions**:** Our refined CG model demonstrates excellent performance in predicting near-native structures of complex DNAs, including three-way and four-way junctions. The structure predictions align well with state-of-the-art models such as 3dRNA/DNA and AlphaFold3, confirming the ability of the model to capture the intricate conformational details of DNA with multi-way junctions with high accuracy.

2. Reliable Stability Profiling: The model provides consistent and accurate predictions of the stability of a wide range of DNA sequences and structures. Notably, it successfully captures the effects of both monovalent and divalent ions on DNA junction stability, showing strong agreement with experimental data. This demonstrates the capacity of the model to simulate biologically relevant conditions and underscores its potential for studying DNA stability under physiological ionic conditions.

3. Insight into Thermally Unfolding Pathways: Through the analysis of thermally unfolding behaviors, we revealed that the stability of multi-way junctions is governed by the relative stabilities of their intermediate states. This finding parallels the behavior of DNA with multi-way junctions and provides deeper insights into the thermodynamic principles that drive DNA unfolding processes. Our study highlights the major unfolding pathways of these structures and offers a detailed understanding of their transition states.

In spite of these strengths, there are several areas for further development. First, the model does not yet account for non-canonical base pairings (e.g., A-A, A-G) [68,69], which could play an important role in the stability and structural features of loops, particularly in internal and DNA triple helix [70,71]. Second, although the model incorporates the effects of monovalent and divalent ions through the CC theory and TBI model, it may not fully capture the stabilizing influence of divalent ions, such as $Mg^{2+}$, which are known to be critical for stabilizing complex DNA structures like G-quadruplex [72,73]. Third, testing the model against a broader range of complex DNA structures, including G-quadruplexes, would further refine its predictive capabilities. Finally, enhancing the accuracy of structure predictions could be achieved by improving the scoring function, which currently limits the full predictive potential of our present model.

Despite these limitations, our refined CG model provides a powerful and reliable framework for studying the 3D structures and stability of complex DNA configurations in the presence of physiologically relevant ions. The insights gained from analyzing the thermally unfolding pathways of DNA junctions contribute to a deeper understanding of DNA stability and the mechanisms underlying their biological functions, laying a solid foundation for future studies in the field of DNA biophysics.

## Acknowledgements

We are grateful to Profs Yang Yu (Guizhou Medical University) and Qiude Li (Guizhou Medical University) for valuable discussions. The numerical calculations in this work were performed on the super computing system in the Super Computing Center of Guizhou Medical University.

## Author Contributions

Y. Z. S. and X. W. designed the research; X. W. performed the simulations; Y. Z. S. and X. W. analyzed the data; X. W. and Y. Z. S. wrote the manuscript. All authors discussed the results and reviewed the manuscript.

## Declaration of Interest

The authors declare no competing interests.

## Supporting Material

Supporting material are available at XXX.

## Funding

This work was supported by grants from the National Science Foundation of China (11605125)

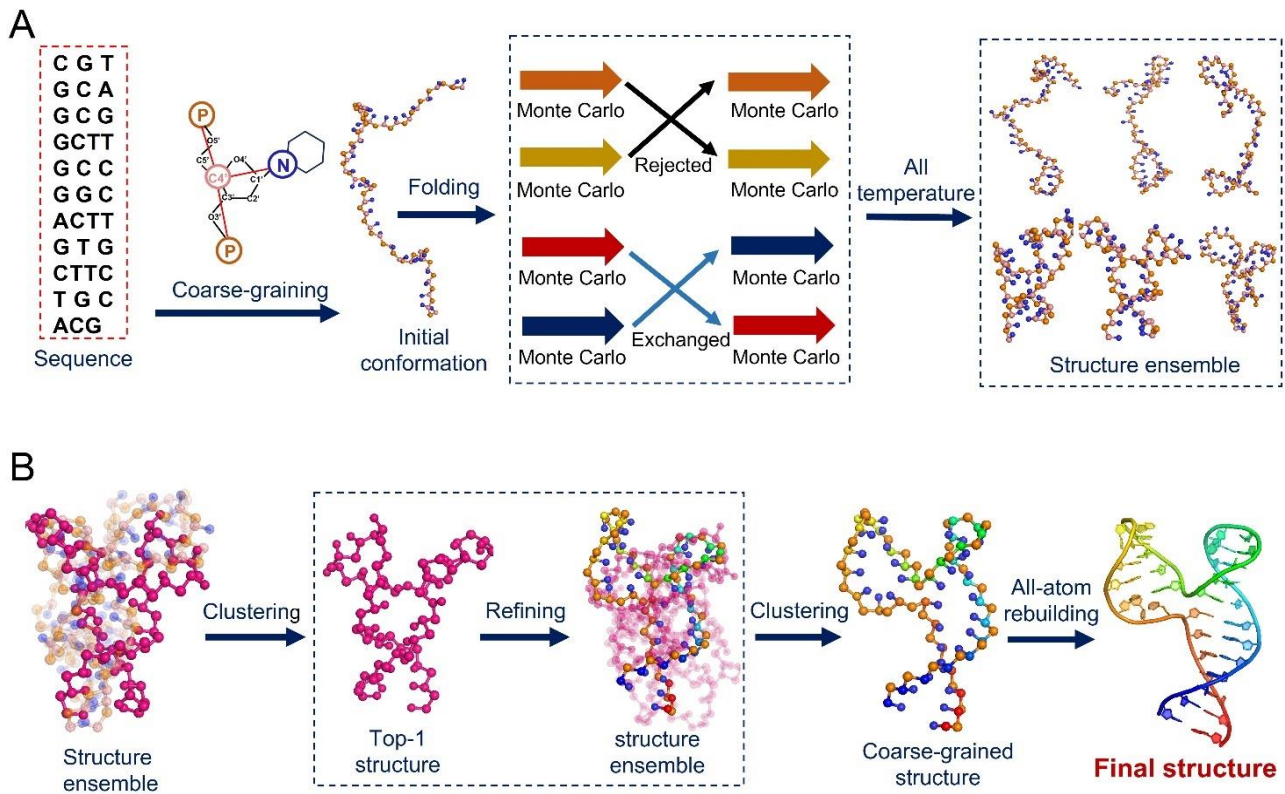

**Figure 1. Schematic representation of the coarse-grained (CG) modeling framework for DNA. (A)** The folding progress of DNA is based solely on sequence information, incorporating the coarse-grained representation of DNA, the initial conformation, REMC simulations, and a structure ensemble derived from ten different temperature replicas to explore conformational diversity. **(B)** Refinement of the DNA structure, beginning with the identification of the most probable conformation through clustering of low-energy conformation, followed by the rebuilding of an all-atom model for detailed structural analysis.

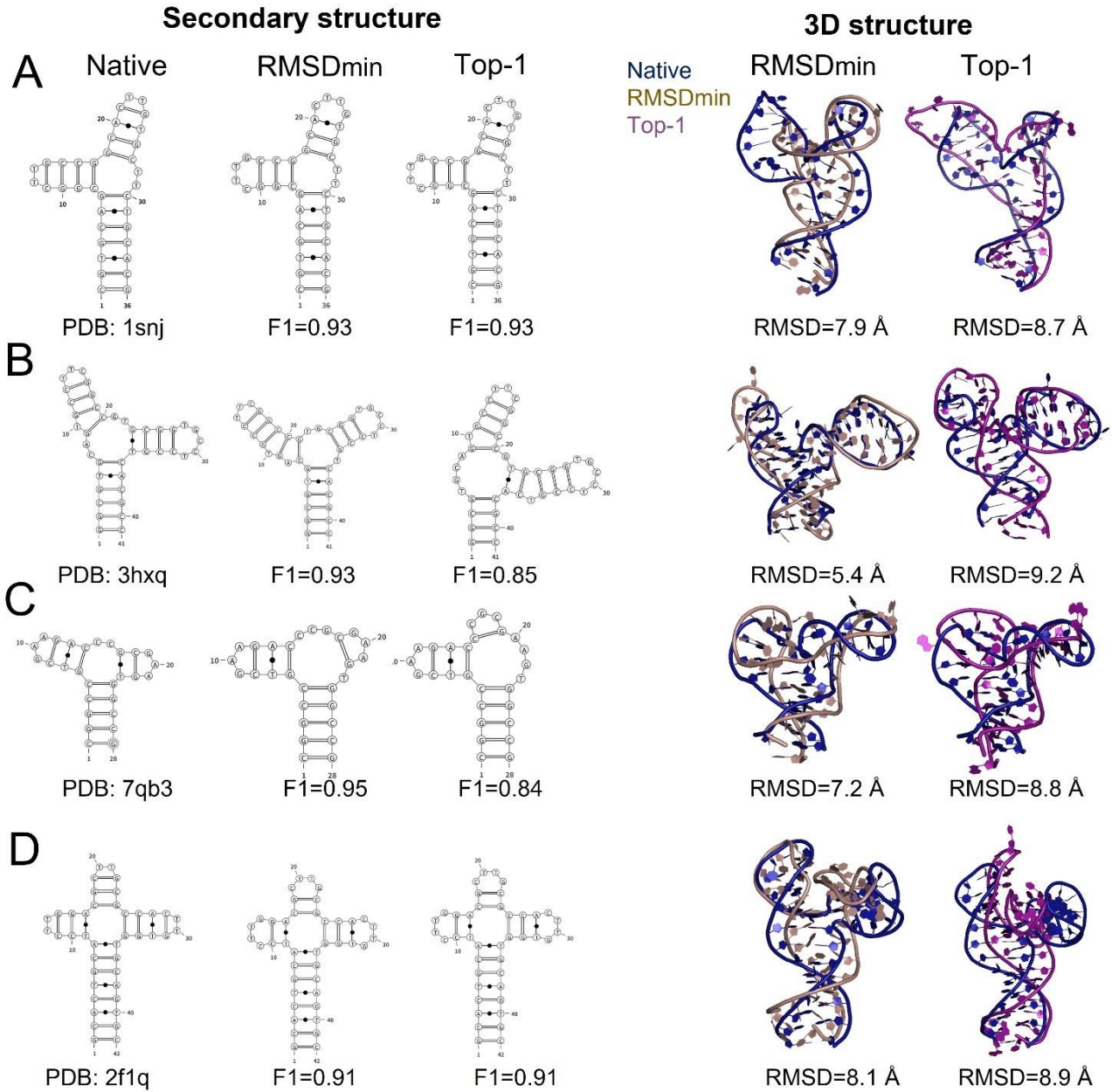

**Figure 2. Comparison of predicted secondary and 3D structures of DNA with RMSD$_{min}$ and top-1 RMSD using our CG model. (A-C)** F1 score and RMSD values for the minimum RMSD and top-1 RMSD structures of DNA with three-way junctions. **(D)** F1 score and RMSD values for the minimum RMSD and top-1 RMSD structures of DNA with four-way junctions. The predicted 3D structures with minimum RMSD (brown) and top-1 RMSD (pink) are overlaid with their respective native structures (blue). Secondary structures were visualized using VARNA [74], and 3D structures were rendered with PyMOL [75].

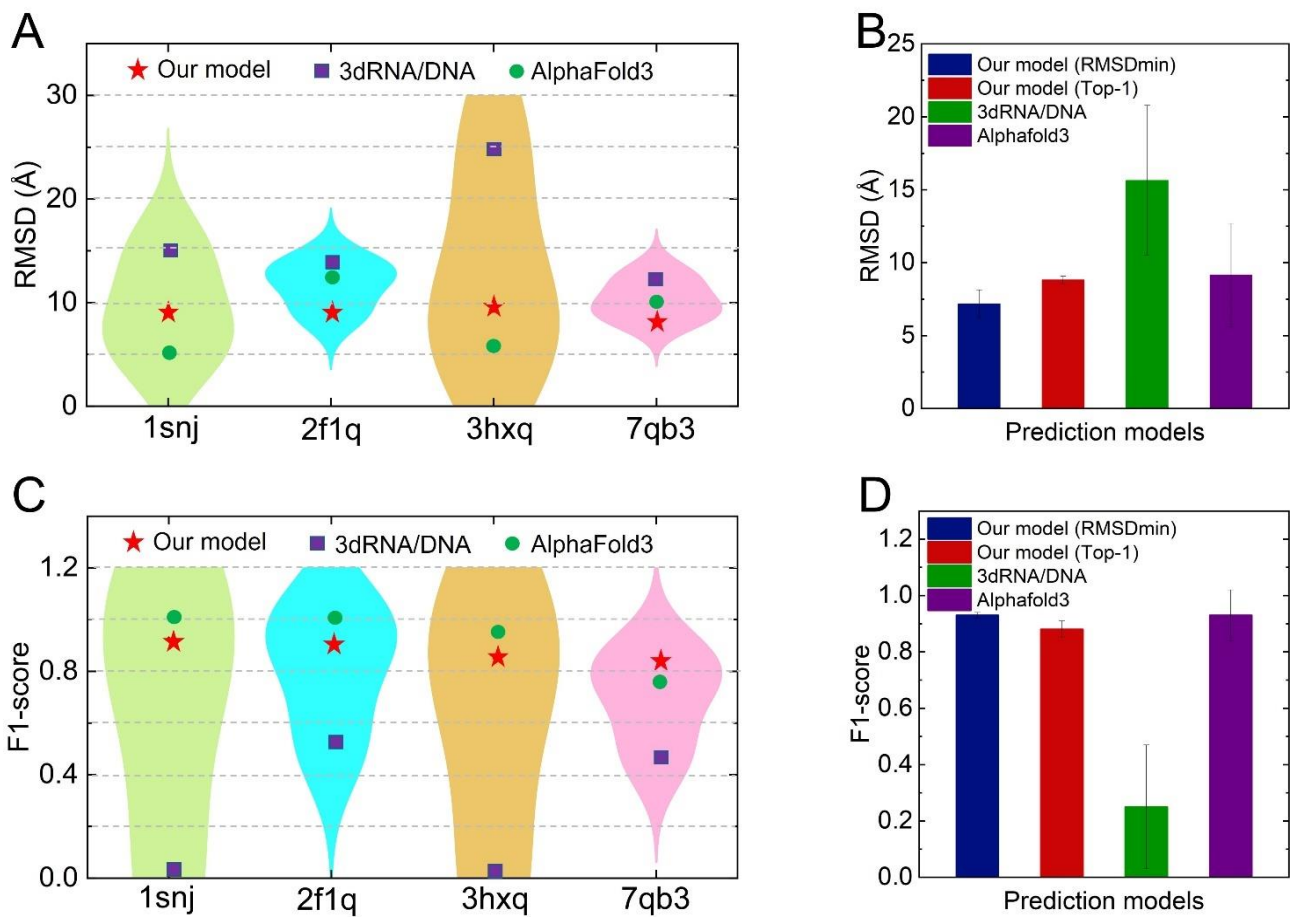

**Figure 3. Comparison of prediction performance between our CG model and other models, including 3dRNA/DNA and AlphaFold3, for four DNA molecules with multi-way junctions.** **(A)** RMSD values for the predictions of four DNA molecules with multi-way junctions using our CG model, 3dRNA/DNA, and AlphaFold3. **(B)** Average RMSD values from our CG model and the other two models for the same four DNA molecules. **(C)** F1 scores for the predictions of the four DNA molecules with multi-way junctions by our CG model, 3dRNA/DNA, and AlphaFold3. **(D)** Average F1 scores from our CG model and the other two models for these DNA structures. Error bars represent the standard deviation of the RMSD and F1 score values.

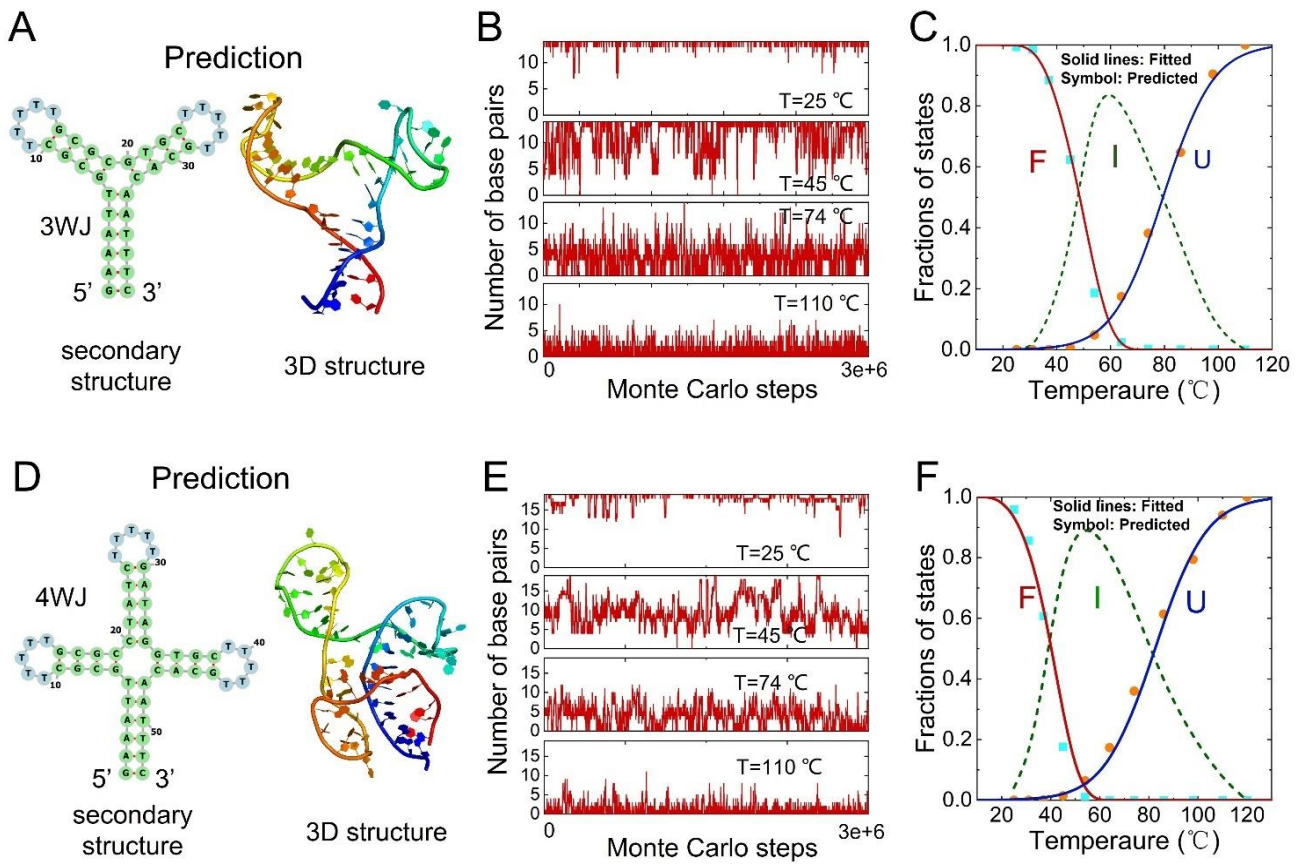

**Figure 4. Predicted melting behavior of DNA with 3-way and 4-way junctions using our CG model.** **(A)** Predicted secondary and 3D structures of the 3-way junction (3WJ) by our CG model. **(B)** Time evolution of the base-pair fractions for 3WJ at different temperatures (110°C, 74°C, 45°C, 25°C, from bottom to top). **(C)** Temperature-dependent fractions of the folded (F, green) and unfolded (U, red) states for 3WJ, simulated in a 1 M NaCl solution. **(D)** Secondary and 3D structures of the 4-way junction (4WJ) as predicted by our model. **(E)** Time evolution of base-pair fractions for 4WJ at temperatures ranging from 110°C to 25°C (bottom to top). **(F)** Fractions of folded (F, green) and unfolded (U, red) states for 4WJ as a function of temperature, also in a 1 M NaCl solution.

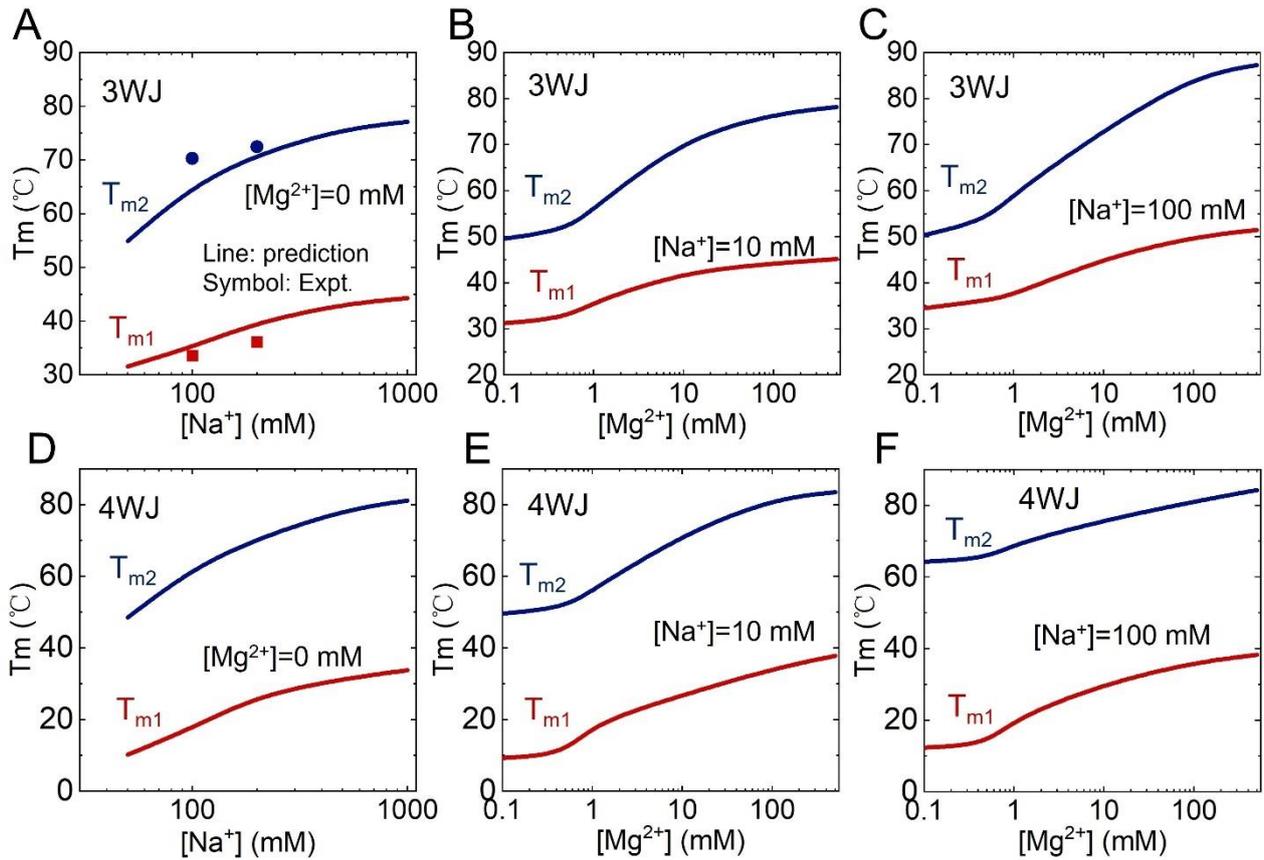

**Figure 5. Comparison of model predictions (lines) with experimental data (symbols) for 3WJ and 4WJ. (A)** Melting temperatures of the F→I ($T_{m1}$) and I→U ($T_{m2}$) transitions for 3WJ as a function of [Na$^+$]. **(B, C)** Melting temperature ($T_{m1}$ and $T_{m2}$) for 3WJ as a function of [Na$^+$] and [Mg$^{2+}$], with [Na$^+$] set at 10 mM (B) and 100 mM (C). **(D)** Melting temperature for 4WJ of the F→I ($T_{m1}$) and I→U ($T_{m2}$) transitions as a function of [Na$^+$]. **(E, F)** Melting temperature ($T_{m1}$ and $T_{m2}$) for 4WJ as a function of [Na$^+$] and [Mg$^{2+}$], with [Na$^+$] set at 10 mM (E) and 100 mM (F).

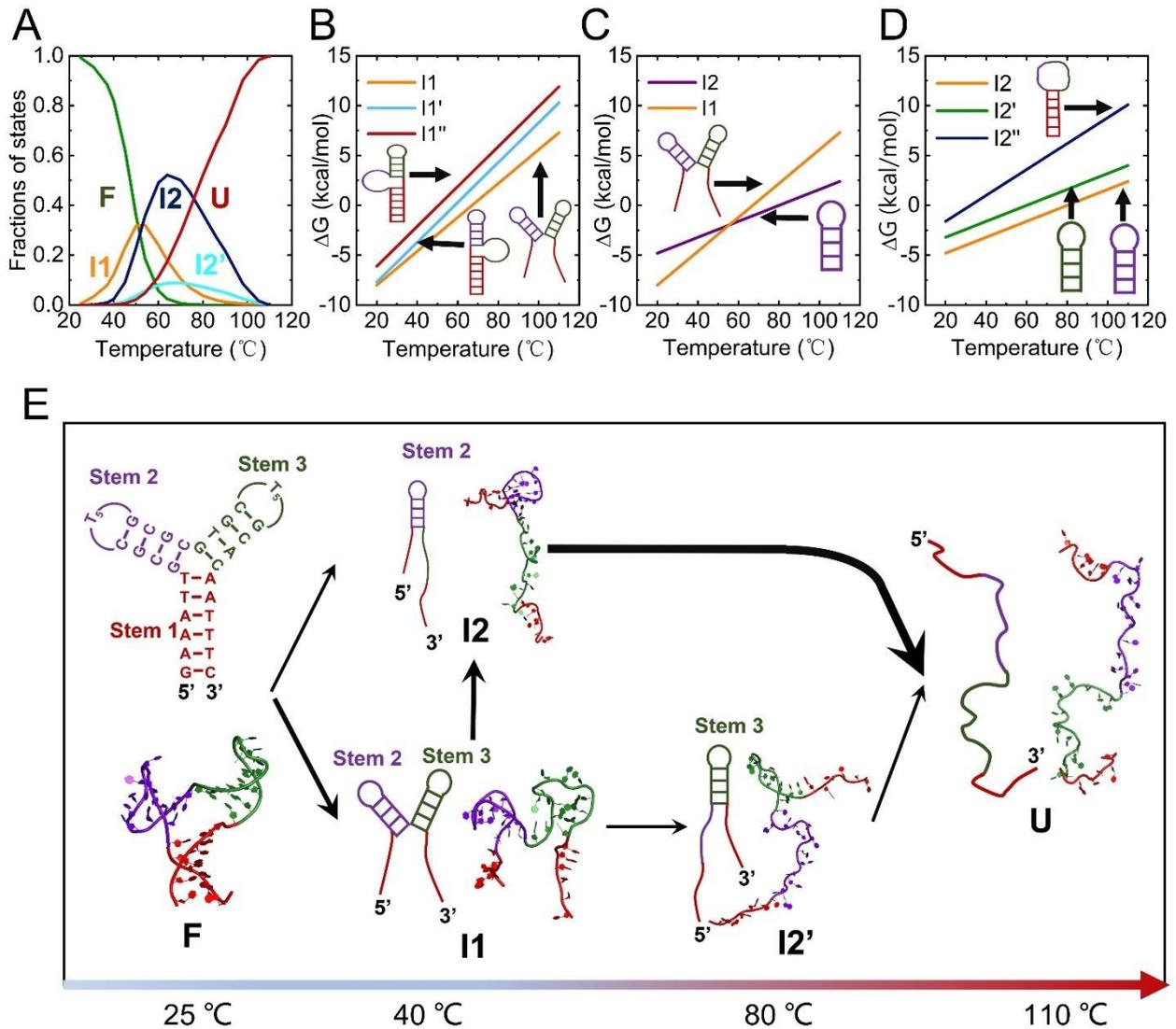

**Figure 6. Predicted thermally unfolding pathway of 3WJ at 1 M [Na$^+$] using our CG model. (A)** Temperature-dependent distributions of the DNA states during the unfolding of 3WJ, showing the fractions of fully folded (F), intermediate (I1, I2, I2′), and fully unfolded (U) states. Here, F represents the fully folded DNA, I1 corresponds to the intermediate with Stems 2 and 3 melted, I2 and I2′ represent the hairpin intermediates with Stem 2 and Stem 3 melted, respectively, and U denotes the fully unfolded DNA. **(B-D)** Temperature dependence of the free energies for the intermediate states (I1, I1′, I1″, I2, I2′, and I2‴) as calculated by the model. **(E)** Schematic representation of the structural transitions along the unfolding pathway inferred from the state fractions shown in panel (A).

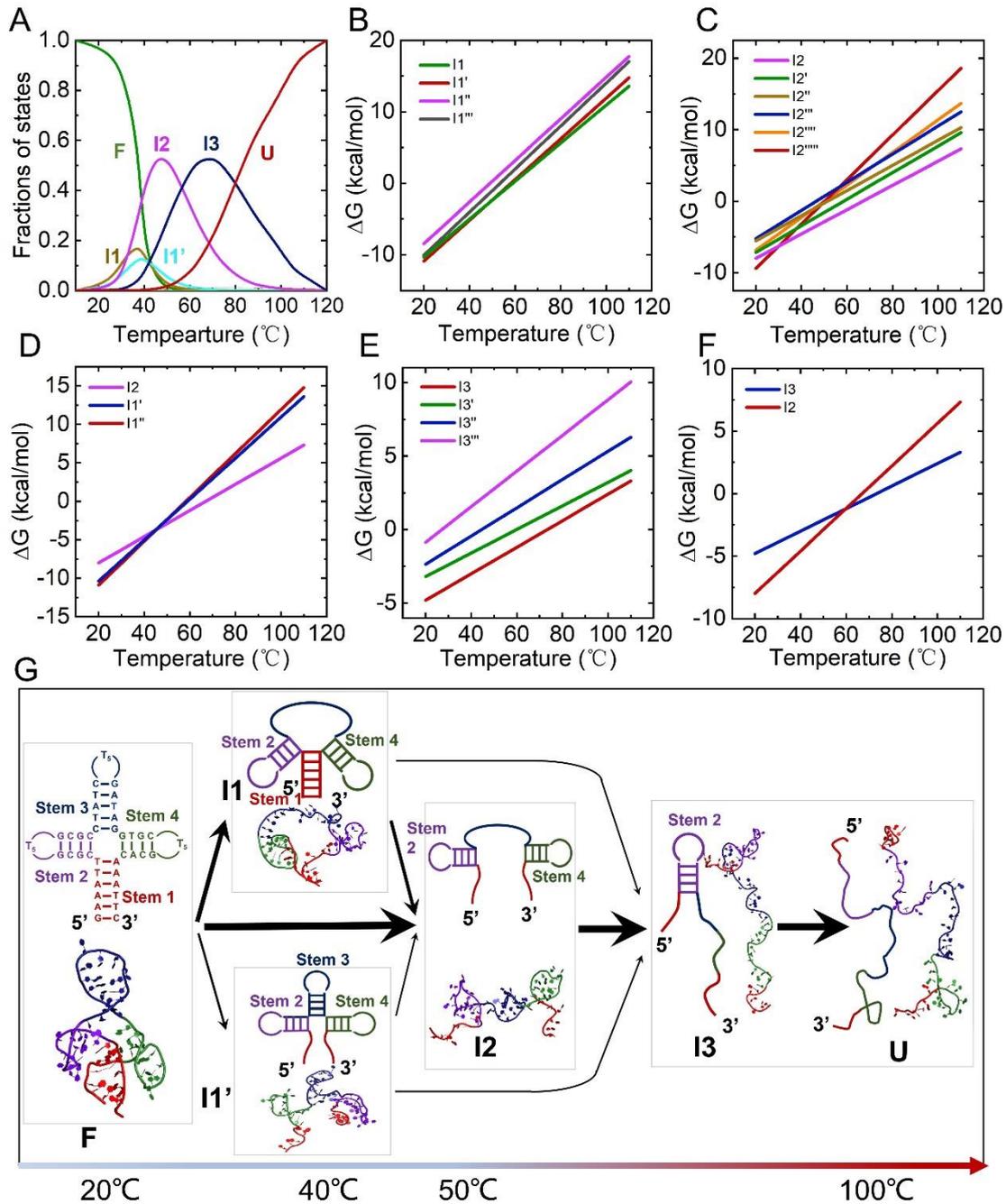

**Figure 7. Predicted unfolding pathway of 4WJ at 1 M [Na$^+$] according to our CG model. (A)** Temperature-dependent fractions of the folded state (F), intermediate states (I1, I1′, I2, I3), and unfolded state (U) during the thermal unfolding pathway of 4WJ at 1 M [Na$^+$]. F, I1, I1′, I2, I3, and U represent the fully folded DNA, the three-way junction intermediates with Stems 1, 2, and 4, and with Stems 2, 3, and 4, the two-way junction with Stems 2 and 4, the hairpin intermediate with Stem 2, and the fully unfolded DNA, respectively. **(C-F)** Free energies of the I1, I1′, I1″, I1‴, I2, I2′, I2″, I2‴, I3, I3′, I3″, and I3‴ states as functions of temperature. **(G)** Schematic representation of DNA structural transitions along the unfolding pathway, inferred from the state fractions in panel (A).

Table 1. The predicted melting temperature ($T_{m1}$[a] and $T_{m2}$[b]) of 6 DNAs by our present model.

| DNAs | Type | Refs | Length (nt) | Predicted $T_{m1}$/$T_{m2}$ (℃) | Expt. $T_{m1}$/$T_{m2}$ (℃) |
|---|---|---|---|---|---|
| 3WJ | Three-way | [64] | 37 | 35.2[c]/65.2[c] | 33.6[c]/70.3[c] |
| L-3WJ | Three-way | [64] | 37 | 37.7[c]/83.8[c] | 35.8[c]/87.6[c] |
| R-3WJ | Three-way | [64] | 37 | 50.6[c] (55.7[d])/74.3[c] | 54.6[d]/77.9[c] |
| 4WJ | Four-way | [64] | 53 | 30.2[c]/73.2[c] | 51.2[e]/65.0[e] |
| L-4WJ | Four-way | [64] | 52 | 31.6[c]/83.5[c] | 52.4[f]/87.9[f] |
| R-4WJ | Four-way | [64] | 52 | 32.9[c]/82.6[c] | 52.8[g]/72.9[g] |

[a] $T_{m1}$ and [b] $T_{m2}$ are the melting temperatures for the transitions from folded state to intermediate state and from intermediate state to unfolded state, respectively. [c] The melting temperature $T_{m1}$ (3 stems transformed into 2 stems) and $T_{m2}$ (1 stem transformed into 0 stem). [d] The melting temperature $T_{m1}$ (3 stems transformed into 1 stem). [e] The melting temperature $T_{m1}$ (4 stems transformed into 1.5 stem) and $T_{m2}$ (1.5 stem transformed into 0 stem). [f] The melting temperature $T_{m1}$ (4 stems transformed into 1.5 stem) and $T_{m2}$ (1 stem transformed into 0 stem). [g] The melting temperature $T_{m1}$ (4 stems transformed into ~1.5 stem) and $T_{m2}$ (~1.5 stem transformed into 0 stem).